\documentclass[prd, aps, nofootinbib,preprintnumbers, showpacs,superscriptaddress,twocolumn]{revtex4}

\usepackage{latexsym}
\usepackage{amssymb}
\usepackage{amsfonts}
\usepackage{amsmath}
\usepackage{bm}
\usepackage[dvips]{graphicx}
\usepackage{color}
\usepackage{times}
\usepackage{units}

\newcommand{\cf}{{cf.}~}
\newcommand{\ie}{{i.e.}~}
\newcommand{\eg}{{e.g.,}~}

\newcommand{\code}[1]{\texttt{#1}}

\allowdisplaybreaks

\begin{document}

\title{Conformal and covariant formulation of the Z4 system with
  constraint-violation damping}

\date{\today}
\label{firstpage}

\author{Daniela Alic}
\affiliation{
 Max-Planck-Institut f\"ur Gravitationsphysik,
 Albert-Einstein-Institut,
  Potsdam-Golm, Germany
}

\author{Carles Bona-Casas}
\affiliation{
Institute for Applied Computation with Community Code
(IAC$^{\,3}$)\\Departament de Fisica, Universitat de les Illes
Balears, 07122 Palma de Mallorca, Spain.
}
\affiliation{
Instituut voor Informatica,
 Universiteit van Amsterdam, 1098 XH Amsterdam, The Netherlands
}

\author{Carles Bona}
\affiliation{
Institute for Applied Computation with Community Code
(IAC$^{\,3}$)\\Departament de Fisica, Universitat de les Illes
Balears, 07122 Palma de Mallorca, Spain.
}

\author{Luciano Rezzolla}
\affiliation{
 Max-Planck-Institut f\"ur Gravitationsphysik,
 Albert-Einstein-Institut,
 Potsdam-Golm, Germany
}
\affiliation{
  Department of Physics and Astronomy,
 Louisiana State University,
  Baton Rouge, Louisiana, USA
}

\author{Carlos Palenzuela}
\affiliation{
Canadian Institute for Theoretical Astrophysics,
 Toronto, Ontario M5S 3H8, Canada
}
\affiliation{
  Department of Physics and Astronomy,
 Louisiana State University,
  Baton Rouge,  Louisiana, USA
}

\begin{abstract}
We present a new formulation of the Einstein equations based on a
conformal and traceless decomposition of the covariant form of the Z4
system. This formulation combines the advantages of a conformal
decomposition, such as the one used in the BSSNOK formulation (\ie
well-tested hyperbolic gauges, no need for excision, robustness to
imperfect boundary conditions) with the advantages of a
constraint-damped formulation, such as the generalized harmonic one
(\ie exponential decay of constraint violations when these are
produced). We validate the new set of equations through standard tests
and by evolving binary black hole systems. Overall, the new
conformal formulation leads to a better behavior of the constraint
equations and a rapid suppression of the violations when they
occur. The changes necessary to implement the new conformal
formulation in standard BSSNOK codes are very small as are the
additional computational costs.
\end{abstract}

\maketitle


\section{Introduction}


Numerical relativity has seen, over the last few years, a truly
remarkable development. Starting from the first simulations showing
that black-hole binaries could be evolved for a few
orbits~\cite{Pretorius:2005gq, Campanelli:2005dd, Baker:2005vv}, or
that black holes could be produced from unstable stellar
configurations using simple gauges and without
excision~\cite{Baiotti06}, new results have been obtained steadily. As
a result, it is now possible to simulate binary black
holes~\cite{Chu:2009md} and binary neutron stars~\cite{Baiotti08}
accurately for dozens of orbits, from the weak-field inspiral, down to
the final black-hole ringdown (see
also~\cite{Centrella:2010,Duez:2009yz} for recent reviews on binary
black holes and neutron stars, respectively). In addition, the
progress in numerical relativity has also been accompanied by a
comparable progress of analytical approximation techniques, which have
been shown to be able to reproduce the numerical results to very high
precision both for binary black
holes~\cite{Damour:2009kr,Buonanno:2009qa} and for binary neutron
stars~\cite{Baiotti:2010}. Finally, numerical simulations have now
investigated scenarios never considered before and that could lead to
a new and deeper understanding of the astrophysics of compact
objects~\cite{Palenzuela:2010,Rezzolla:2011}.

There are several reasons behind this rapid progress, and the use of
more accurate numerical techniques and the availability of larger
computational facilities are certainly among the most important
ones. None of these, however, would be useful without the use of
formulations of the Einstein equations that are well-suited for
numerical evolutions. Most of the present three-dimensional (3D)
numerical-relativity codes implement either one of the two formulations
discussed below. The first and most popular one is the conformal and
traceless reformulation of the $3+1$ ADM equations~\cite{York79},
which is also known as the BSSNOK (or BSSN)
formulation~\cite{Nakamura87, Shibata95, Baumgarte99}. The second
formulation is instead based on the use of a fully 4D form of the
Einstein equations in coordinates that resemble the harmonic ones and
is therefore known as the Generalized-Harmonic formulation
(GH)~\cite{Friedrich85}.

There are several differences between these two formulations, each
having its own advantages and disadvantages. One of the main
advantages of BSSNOK is that, being based on a conformal
decomposition, it can separate potential singular terms in the
conformal factor. In addition, it can count on well-tested and robust
gauge conditions, such as the singularity-avoiding slicing conditions
of the $1+\log$ family~\cite{Bona94b}. Similarly, the spatial gauges
can rely on the hyperbolic Gamma-driver condition for the shift
vector~\cite{Alcubierre02a} (or some recent variants for unequal-mass
binaries~\cite{Mueller:2010bu, Schnetter:2010cz, Alic:2010}), which
removes to a large extent, the gauge dynamics near the compact
objects. When combined, these two gauge choices eliminate the
need to excise a region of the computation domain inside the apparent
horizon, greatly simplifying the numerical infrastructure. Finally, the
use of the momentum constraint equations (but not of the energy
constraint) in the evolution of the dynamical variables, 
which is crucial for ensuring strong hyperbolicity, provides
BSSNOK with a certain ``forgiveness'', so that the violation of the
constraints does not grow rapidly, even when boundary conditions which
are constraint-violating are used near the strong-field region.

In contrast, the GH formulation uses a generalized harmonic gauge
which cannot deal with the physical singularity inside the apparent
horizon. As a result, at least for the gauges considered so far (see
also ~\cite{Pretorius:2004jg, Lindblom:2009}), it requires the use of
excision and thus of numerical techniques that are devised for
handling a special region of the computational
domain~\cite{Szilagyi:2006qy}. To its advantage, however, the GH
formulation leads to a set of equations whose principal parts are wave
equations and thus with very well-known mathematical properties. In
addition, the use of damping terms allows for the dynamical control of
the constraint violations and thus for a powerful way of reducing them
when necessary. Of course, a solution with smaller constraint
violations will intrinsically be a more accurate solution to the
Einstein equations.

Clearly, it would be useful to employ a formulation of the Einstein
equations that combines the best of both worlds and thus that has the
robustness and gauge conditions of the BSSNOK formulation but, at the
same time, has well-defined mathematical properties and the
possibility of dynamically controlling the constraint violations as
the GH formulation. As we will show, these properties are met by a new
conformal and covariant formulation of the Z4 system with
constraint-violation damping. This is obtained by starting from the
fully covariant Z4 formulation~\cite{Bona:2003fj} and by performing a
conformal decomposition which includes \textit{all} the nonprincipal
terms coming from the covariant form of the equations. In addition,
damping terms are included for controlling the constraints in the
spirit of the GH formulation. We will refer to this new formulation as
the conformal and covariant Z4 system, \ie CCZ4, and present tests of
its behavior by considering evolutions in vacuum of gauge waves in 1D
and isolated and binary black holes in 3D.

It should be remarked that this is not the first time that a conformal
decomposition of the Z4 system has been proposed and indeed a very
interesting attempt has been made in Ref.~\cite{Bernuzzi:2009ex},
where it was named Z4c. Although the tests presented in
Ref.~\cite{Bernuzzi:2009ex} were performed in spherical symmetry, they
already highlighted the potential of a conformal formulation of the Z4
system, especially in the presence of matter (see
also~\cite{Ruiz:2010qj,Weyhausen:2011cg}). Unfortunately, we were not
able to obtain equally good results when evolving the formulation of
Ref.~\cite{Bernuzzi:2009ex} in vacuum and in 3D; at the same time, we
did not find that our CCZ4 formulation is more sensitive to boundary
problems than the BSSNOK one (this was a point raised in
Ref.~\cite{Bernuzzi:2009ex}).

The structure of the paper is as follows. In Sec.~\ref{the system},
we derive the full set of the CCZ4 equations starting from the
covariant form of the Z4 system. In Sec.~\ref{numerical results} we
introduce the details of the numerical infrastructure and present a
numerical comparison between the CCZ4 and the BSSNOK systems for a
gauge-wave test and for binary black-hole simulations. Finally, the
conclusions are summarized in Sec.~\ref{conclusions}.


\section{The conformal covariant Z4 system}
\label{the system}

The Z4 formulation was introduced as a covariant extension of the
Einstein equations~\cite{Bona:2003fj}, where the original elliptic
constraints are converted into algebraic conditions for a new
four-vector $Z_{\mu}$. 
This formulation can be derived from the covariant
Lagrangian
\begin{equation}\label{action}
{\cal L} = g^{\mu\nu}~[R_{\mu\nu} + 2~\nabla_{\mu} Z_{\nu}]\,,
\end{equation}
by means of a Palatini-type variational principle~\cite{Bona:2010is}.
The vector $Z_{\mu}$ measures the deviation from the Einstein field
equations. The algebraic constraints $Z_{\mu}=0$ amount therefore to
the fulfilling of the standard energy-momentum constraints. In order
to control these constraints, the original system was supplemented
with damping terms such that the true Einstein solutions (\ie the ones
satisfying the constraints) become an attractor of the enlarged set of
solutions of the Z4 system~\cite{Gundlach2005:constraint-damping}. 
The Z4 damped formalism can be written in covariant form as
\begin{eqnarray}\label{Z4_covariant}
     R_{\mu\nu} &+& \nabla_{\mu} Z_{\nu} + \nabla_{\nu} Z_{\mu}  +
     \kappa_1 [n_{\mu} Z_{\nu} + n_{\nu} Z_{\mu} \nonumber \\ &-& (1 + \kappa_2)
     g_{\mu\nu} n_{\sigma} Z^{\sigma}] = 8 \pi (T_{\mu\nu} - \tfrac{1}{2} g_{\mu\nu} T)\,,
\end{eqnarray}
where $n_{\mu}$ is the unit normal to the time slicing, $T_{\mu\nu}$
the stress-energy tensor and $T$ its trace, \ie $T\equiv g_{\mu\nu}
T^{\mu\nu}$. The (constant) coefficients $\kappa_i$ are free
parameters related to the characteristic time of the exponential
damping of constraint violations. Assuming energy-momentum tensor
conservation, the Bianchi identities lead to the
constraint-propagation system
\begin{equation}
\label{constraints}
     \nabla^{\nu} \nabla_{\nu} Z_{\mu} + R_{\mu\nu} Z^{\nu} =
   - \kappa_1 \nabla^{\nu} [n_{\mu} Z_{\nu} + n_{\nu} Z_{\mu} + \kappa_2 g_{\mu\nu} n_{\sigma} Z^{\sigma}]\,.
\end{equation}
It has been shown in Ref.~\cite{Gundlach2005:constraint-damping} that
all the constraint-related modes are damped when
\begin{equation}\label{damping pars}
    \kappa_1 > 0 \qquad \kappa_2 > -1\,.
\end{equation}

The Z4 formulation can be rewritten as a Cauchy problem by performing
the $3+1$ decomposition of the spacetime, in which the line element
reads
\begin{equation}\label{3+1metric}
    ds^2 = - \alpha^2 dt^2 + \gamma_{ij}\; (dx^i+\beta^i dt)\,
    (dx^j+\beta^j dt)\,,
\end{equation}
where $\alpha$ is the lapse function, $\beta^i$ is the shift vector
and $\gamma_{ij}$ the intrinsic metric of the constant-time
slices. The Einstein equations within this decomposition lead to
the well-known ADM system~\cite{York79}, which is usually cast as a
system of evolution equations for the extrinsic curvature $K_{ij}$ and
the three-metric $\gamma_{ij}$, plus four elliptic equations for the
energy (or Hamiltonian) and the momentum constraints, involving space
derivatives of the dynamical fields $\gamma_{ij}$ and $K_{ij}$.  In
the Z4 formulation, the energy-momentum constraints become evolution
equations for $Z_{\mu}$, modifying the principal part of the ADM
system and converting it from weakly to strongly
hyperbolic~\cite{Bona:2003qn}. The $3+1$ decomposition of the Z4
formulation including the damping terms reads
\begin{widetext}
\begin{eqnarray}
\label{Z4-dtgamma}
  (\partial_t &-&{\cal L}_{\beta})~ \gamma_{ij}
  = - {2\,\alpha}\,K_{ij} \,,\\
\label{Z4-dtK}
   (\partial_t &-& {\cal L}_{\beta})~K_{ij} = -\nabla_i\alpha_j
    + \alpha\;   \left[\,R_{ij} + \nabla_i Z_j+\nabla_j Z_i
    - 2\,{K_i}^{l}\, K_{lj}+(K-2\Theta)\,K_{ij}
    -\kappa_1(1+\kappa_2)\,\Theta\,\gamma_{ij}\,\right] \nonumber \\
    &&\hskip 1.7cm -  8\pi\alpha\, \left[\,S_{ij}-\frac{1}{2}\,(S - \tau)\,\gamma_{ij}\,\right] \,,\\
\label{Z4-dtTheta} (\partial_t &-& {\cal L}_{\beta})~\Theta =
\frac{\alpha}{2}\; \left[\, R + 2\, \nabla_j Z^j + (K -
2\Theta)\, K
 - K^{ij}\, K_{ij} - 2 \frac{Z^j {\alpha}_j}{\alpha}
-2\, \kappa_1 (2+\kappa_2)\,\Theta- 16\pi\,\tau\,\right] \,,\\
\label{Z4-dtZ}
 (\partial_t &-& {\cal L}_{\beta})~Z_i =
 \alpha\, [\,\nabla_j\,({K_i}^j
  -{\delta_i}^j K) + \partial_i \Theta
  - 2\, {K_i}^j\, Z_j  -  \Theta\, \frac{{\alpha}_i}{\alpha}
  -\kappa_1 Z_i- 8\pi\,S_i\,]\,,
\end{eqnarray}
\end{widetext}
where ${\cal L}_{\beta}$ is the Lie derivative along the shift vector
$\vec{\beta}$, $\Theta$ is the projection of the Z4 four-vector along
the normal direction, $\Theta \equiv n_{\mu} Z^{\mu} = \alpha Z^0$,
and the following definitions apply for matter-related quantities $\tau
\equiv n_{\mu} n_{\nu} T^{\mu\nu}$, $S_i \equiv n_{\nu} T^{\nu}_{~i}$,
$S_{ij} \equiv T_{ij}$.

Equations~\eqref{Z4-dtgamma}--\eqref{Z4-dtZ} must be complemented with
suitable gauge conditions that determine the system of coordinates
used during the evolution. Of all the possible options, the most
interesting ones are those which preserve the hyperbolicity of the
full evolution system, such as the $1+\log$ family and the
Gamma-driver shift condition.

As a first step towards deriving the CCZ4 formulation, we express the
metric $\gamma_{ij}$ in terms of a conformal metric
${\tilde{\gamma}}_{ij} = \phi^2 \gamma_{ij} $ with unit determinant
$\phi = ({\rm det} (\gamma_{ij}))^{-1/6}$, while the extrinsic
curvature $K_{ij}$ is decomposed into its trace $K \equiv K_{ij} \gamma^{ij}$ and in its trace-free components
\begin{equation}
\label{tracelessK}
{\tilde{A}}_{ij} = \phi^2\;(K_{ij}- \frac{1}{3} K \gamma_{ij})\,.
\end{equation}
This allows us to write the three-dimensional Ricci tensor as $R_{ij}
= \tilde R_{ij} + \tilde R^{\phi}_{ij}$, thus splitting it into a part
containing conformal terms and another one containing space
derivatives of the conformal metric
\begin{widetext}
\begin{eqnarray}
\tilde R_{ij} &=& -\frac{1}{2} \tilde \gamma^{lm} \partial_l \partial_m \tilde \gamma_{ij}
  + \tilde \gamma_{k(i} \partial_{j)} \tilde \Gamma^k
  + \tilde \Gamma^k \tilde \Gamma_{(ij)k}
  + \tilde \gamma^{lm} \left[2 {\tilde \Gamma^k}_{l(i}
\tilde \Gamma_{j)km} + {\tilde \Gamma^k}_{im} \tilde \Gamma_{kj\,l}\right]\,, \\
\tilde R^{\phi}_{ij} &=& \frac{1}{\phi^2}\left[\phi \left(\tilde \nabla_i \tilde
\nabla_j \phi + \tilde \gamma_{ij} \tilde \nabla^l \tilde \nabla_l
\phi\right) - 2 \tilde \gamma_{ij} \tilde \nabla^l \phi \tilde \nabla_l
\phi\right]\,,
\end{eqnarray}
\end{widetext}
where
\begin{equation}\label{localGamma}
\tilde \Gamma^i \equiv \tilde \gamma^{jk} \tilde \Gamma^{i}_{jk} =
\tilde \gamma^{ij} \tilde \gamma^{kl} \partial_l \tilde
\gamma_{jk}\,.
\end{equation}
The conformal and covariant Z4 formulation (CCZ4) is thus given by the
following system of evolution equations
\begin{widetext}
\begin{eqnarray}
\partial_t\tilde\gamma_{ij} &=& - 2\alpha \tilde A^{^{\rm TF}}_{ij}
+ 2\tilde\gamma_{k(i}\partial_{j)}~\beta^k
- \frac{2}{3}\tilde\gamma_{ij}\partial_k~\beta^k
+\beta^k \partial_k \tilde\gamma_{ij} \,,  \label{gamma_eq}\\
\partial_t \tilde A_{ij} &=& \phi^2  \left[-\nabla_i \nabla_j \alpha
+ \alpha \left(R_{ij} + \nabla_i Z_j + \nabla_j Z_i - 8 \pi S_{ij}\right)\right]^{\rm TF}
+ \alpha \tilde A_{ij}\left(K- 2\Theta\right) \nonumber \\
&&
- 2\alpha \tilde A_{il}\tilde A^l_j 
+ 2\tilde A_{k(i}\partial_{j)}~\beta^k
-\frac{2}{3}\tilde A_{ij}\partial_k~\beta^k + \beta^k \partial_k \tilde A_{ij}  \,,  \label{A_eq} \\
\partial_t\phi &=& \frac{1}{3} \alpha \phi K
- \frac{1}{3} \phi \partial_k \beta^k + \beta^k \partial_k \phi \,, \\
\partial_t K &=& - \nabla^i \nabla_i \alpha + \alpha \left(R + 2
  \nabla_iZ^i + K^2 -2 \Theta K \right)
+ \beta^j \partial_j K - 3 \alpha \kappa_1 \left(1 +
\kappa_2\right) \Theta
+ 4 \pi \alpha \left(S - 3 \tau\right) \,, \\
\partial_t \Theta &=& \frac{1}{2} \alpha \left(R + 2 \nabla_i Z^i - \tilde A_{ij} \tilde
A^{ij} + \frac{2}{3} K^2 - 2 \Theta K\right) - Z^i
\partial_i \alpha+ \beta^k \partial_k \Theta
- \alpha \kappa_1 \left(2 + \kappa_2\right) \Theta - 8\pi \alpha\,\tau\,, \\
\partial_t \hat\Gamma^i &=& 2\alpha \left(\tilde\Gamma^i_{jk} \tilde A^{jk}
- 3 \tilde A^{ij} \frac{\partial_j \phi}{\phi} - \frac{2}{3}
\tilde\gamma^{ij} \partial_j K \right)
+2\tilde\gamma^{ki}\left(\alpha \partial_k \Theta - \Theta
\partial_k \alpha
- \frac{2}{3} \alpha K Z_k\right) -  2\tilde A^{ij} \partial_j \alpha \nonumber \\
&& + \tilde\gamma^{kl} \partial_k \partial_l \beta^i
+ \frac{1}{3}\tilde\gamma^{ik}\partial_k\partial_l \beta^l
+ \frac{2}{3} \tilde\Gamma^i \partial_k \beta^k -
\tilde\Gamma^k \partial_k \beta^i 
+ 2 \kappa_3 \left(\frac{2}{3} \tilde\gamma^{ij} Z_j \partial_k \beta^k -
 \tilde\gamma^{jk} Z_j \partial_k \beta^i \right) \nonumber \\
&&+ \beta^k \partial_k \hat\Gamma^i
- 2 \alpha \kappa_1 \tilde \gamma^{ij} Z_j - 16 \pi \alpha
\tilde\gamma^{ij} S_{j}\,, \label{Gamma_eq}\\
\label{1plog}
\partial_t \alpha &=& -2\alpha \left(K-2\Theta\right) + \beta^k\partial_k \alpha \,, \\
\label{gammadriver1}
\partial_t \beta^i &=& f B^i +\beta^k\partial_k\beta^i \,, \\
\label{gammadriver2}
\partial_t B^i &=& \partial_t \hat\Gamma^i - \beta^k \partial_k \hat\Gamma^i
+ \beta^k\partial_k B^i - \eta B^i \,,
\end{eqnarray}
\end{widetext}
where we have defined
\begin{equation}
\label{Gammai}
\hat \Gamma^i \equiv \tilde \Gamma^i + 2 \tilde \gamma^{ij} Z_j\,.
\end{equation}

Note that the choice made with the
definition~\eqref{Gammai} is equivalent, in the ADM context, to adding
the momentum constraint to the right-hand-side of the evolution
equation of $\tilde \Gamma^i$. In the context of the Z4 formulation,
this just amounts to replacing the vector $Z_i$ by the quantities
${\hat \Gamma}^i$ in the set of basic fields to be evolved.

The gauge conditions~\eqref{1plog}--\eqref{gammadriver2} correspond
respectively to the standard ``$1+\log$'' slicing condition and to the
original form of the gamma-driver shift condition, where a generic
gauge parameter $f$ was introduced~\cite{Alcubierre02a}. Note that in
the Z4 formulation there is an additional propagation speed and the
standard BSSNOK choice of $f =3/4$ can then lead to weak hyperbolicity
when the lapse $\alpha$ is close to 1. This is why safer choices,
such as $f=1$, have been proposed in Ref.~\cite{Bernuzzi:2009ex}. In
this paper we use $f=3/4$ to be as close as possible to a
standard BSSNOK formulation, but we also consider how the system of
equations reacts when switching to $f=1$.

We also note that experimentation with black-hole spacetimes and the
emergence of unstable behaviors, has induced us to introduce an extra
parameter, $\kappa_3$, affecting some quadratic terms in the evolution
Eq. \eqref{Gamma_eq} for ${\hat \Gamma}^i$. As discussed before, this
equation corresponds to the evolution of $Z_i$, so this is not just a
gauge choice, but rather an essential ingredient of the Z4
system. Indeed, the covariance inherent to the conformal decomposition
of the Z4 system is broken unless we take $\kappa_3 = 1$. For some of
the tests presented in this paper we retain a fully covariant
formulation (\ie with $\kappa_3 = 1$). However, this is not possible
for black-hole spacetimes, where nonlinear couplings with the damping
terms, which are important for reducing the violations in the
constraints, lead to numerical instabilities. As a result, for
black-hole spacetimes we have resorted to a noncovariant and conformal
formulation of the Z4 system (\ie with $\kappa_3 =1/2$) (see
discussion in Sec.~\ref{Black-Hole Spacetimes} for details). 

A number of remarks are important at this point. First, although
the structure of the CCZ4 formulation is very similar to the BSSNOK
one, there is an important difference in the evolution of the
trace-free variable $\tilde A_{ij}$. In the BSSNOK formulation, in
fact, the Hamiltonian constraint is assumed to be satisfied
\textit{exactly} and thus used to eliminate the Ricci scalar from the
right-hand-side of the evolution equation for $\tilde
A_{ij}$~\cite{Alcubierre02a}. In the CCZ4 system, on the other hand,
the evolution of $\tilde A_{ij}$ follows directly from (the trace-free
part of) the original ADM evolution equation for the extrinsic
curvature $K_{ij}$, plus the extra terms in $Z_i$ and
$\Theta$. Second, the equivalent of the trace of the extrinsic
curvature in BSSNOK formulations is given by
\begin{equation}
\label{BSSN_K}
    K^{^{\rm BSSNOK}} = K - 2\, \Theta\,,
\end{equation}
again because the Hamiltonian constraint is assumed to remove the
Ricci scalar from the evolution equations in the BSSNOK approach. In
the CCZ4 system, we rather use (the trace part of) the ADM evolution
equation for $K_{ij}$, modulo some $Z_i$ and $\Theta$ terms.

A closer look at the resulting CCZ4 system shows that it is not fully
equivalent to the Z4 system, modulo a rearrangement of the dynamical
fields. There are two extra fields which were not present in the Z4
system, namely $\det\,\tilde \gamma_{ij}$ and ${\rm tr}\,\tilde
A_{ij}$. These are not dynamical fields at the continuum level, where
the consistency constraints
\begin{equation}
\label{xtra constraints}
    \det\,\tilde \gamma_{ij} = 1\,,\qquad {\rm tr}\,\tilde A_{ij}=0\,,
\end{equation}
hold by construction. But at the discrete level, these are just
two more constraints, which can be dealt with in many different
ways. For instance:
\begin{itemize}
    \item {\it Constrained approach}. We could enforce (\ref{xtra
      constraints}) at every integration step, by removing the trace
      of $\tilde A_{ij}$ and rescaling $\tilde \gamma_{ij}$ as it is
      usually done in BSSNOK codes~\cite{Alcubierre99d}. The remaining
      dynamical modes have then the same characteristic structure of
      the original Z4 system. This is the safest choice, and we will
      use it in the tests presented in this paper.

    \item {\it Relaxed approach}. We could instead relax (\ref{xtra
      constraints}), enforcing it just on the initial/boundary data.
      In this way the two extra dynamical modes propagate along normal
      lines, as their evolution equations [\ie the trace of
      Eqs. (\ref{gamma_eq})-(\ref{A_eq})] are trivial. Note that in
      this case the trace of the first term in the evolution 
      Eq. (\ref{gamma_eq}) must be removed explicitly to avoid
      any spurious numerical modes by evolving: 
\begin{eqnarray}
 \partial_t\tilde\gamma_{ij} &=& - 2\alpha \left(\tilde A_{ij} -
\frac{1}{3} \tilde \gamma_{ij} \tilde A_{kl} \tilde \gamma^{kl}\right) \nonumber\\
& & + 2\tilde\gamma_{k(i}\partial_{j)}~\beta^k
- \frac{2}{3}\tilde\gamma_{ij}\partial_k~\beta^k
+\beta^k \partial_k \tilde\gamma_{ij}. \nonumber
\end{eqnarray}
      Moreover, in tests like the robust stability or the gauge waves,
      it may be necessary to keep also under control the trace of
      $\tilde A_{ij}$. This can be achieved by adding, for instance, a
      damping term proportional to $\tilde \gamma_{ij}\, {\rm tr}
      \tilde A_{ij}$ to the evolution Eq. (\ref{A_eq}).
\end{itemize}

Finally, the ADM constraints are given by
\begin{eqnarray}
H &=& R - K_{ij} K^{ij} + K^2\,, \\
M_i &=&  \gamma^{jl} (\partial_{l} K_{ij} - \partial_{i} K_{jl} -
\Gamma^{m}_{jl} K_{mi} + \Gamma^{m}_{ji} K_{ml})\,.
\end{eqnarray}
In the results presented below we compute the constraint violations
for both the BSSNOK and CCZ4 systems using the ADM quantities computed
from the evolution variables corresponding to the two systems,
allowing for the correspondence (\ref{BSSN_K}).


\section{Numerical Results}
\label{numerical results}


In this section we validate the robustness and accuracy of the CCZ4
evolution system and compare it against the BSSNOK system in two
different cases: the gauge-waves test and black-hole spacetimes.  In
addition, we have performed several evolutions with the
robust-stability test to ensure that the system is stable to linear
perturbations, recovering the expected results
(see~\cite{Alcubierre:2003pc} for a discussion of this test). 

The numerical setup used in the simulations presented here is the same
one discussed in Ref.~\cite{Pollney:2007ss} and more recently applied
to the \code{Llama} code described in Ref.~\cite{Pollney:2009yz}. The
latter makes use of higher-order finite-difference algorithms
satisfying the summation-by-parts rule (up to 8th order in space) and
a multiblock structure for the outer computational domain. More
specifically, we use a central cubical Cartesian patch containing
multiple levels of adaptive mesh refinement with higher-resolution
boxes. The Cartesian grid is surrounded by $6$ additional patches with
the grid points arranged in a spherical-type geometry, with constant
angular resolution to best match the resolution requirements of
radially outgoing waves. This allows us to move the outer boundary to
a radius where it is causally disconnected from the binary at a tiny
fraction of the computational cost which would be necessary to achieve
the same resolution with a purely Cartesian code. The time evolution
is based on the method-of-lines with a 4th order Runge-Kutta
algorithm.  Our general computational infrastructure is based on the
\texttt{Cactus} framework and we are using packages such as
\texttt{TwoPunctures}~\cite{Ansorg:2004ds},
\texttt{AHFinderDirect}~\cite{Thornburg2003:AH-finding_nourl} and of
\texttt{SummationByParts}~\cite{Diener:2005tn}, which are freely available
and part of the Einstein Toolkit. In addition, our evolutions make use
of the mesh-refinement driver \texttt{Carpet}
\cite{Schnetter-etal-03b}, which implements higher-resolution boxes
with multiple levels of adaptive mesh refinement.


\subsection{Gauge Waves}

A classical test for different formulations of the Einstein equations
is offered by the ``gauge-wave''~\cite{Alcubierre:2003pc}, in which a
fictitious one-dimensional pulse propagating along the $x$-direction
can be simulated by performing a conformal transformation of the
Minkowski metric in the two-dimensional sector spanned by the $(t,x)$
coordinates, namely using the line element
\begin{equation}\label{Minktrans}
    ds^2 = h(x,t)~(-dt^2 + dx^2) + dy^2 + dz^2\,.
\end{equation}
%

\begin{figure}
\begin{center}
\includegraphics[width=8.0cm]{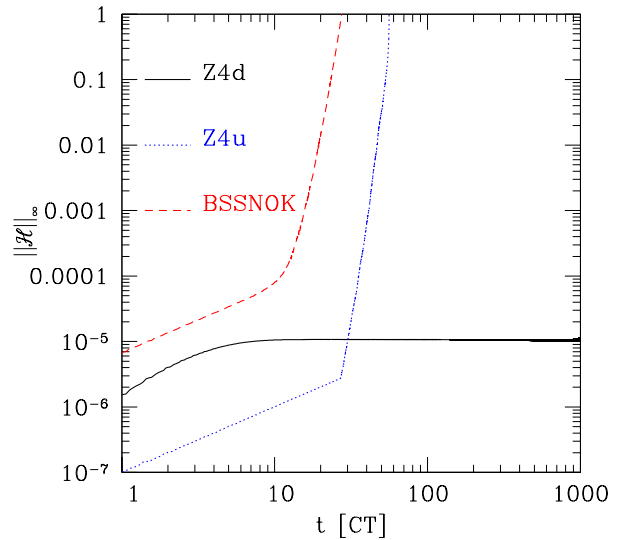}
\caption{L-infinity norm of the Hamiltonian constraint in the
  gauge-wave test, when performed with a CCZ4 formulation with damping
  terms (black solid line), with a CCZ4 formulation without damping
  terms (blue dotted line), or with the BSSNOK formulation (red dashed
  line). Clearly, the Z4u and the BSSNOK formulations are unstable
  (\cf Fig.~5 of Ref.~\cite{Babiuc:2007vr}) and a similar behavior
  will be encountered also in black-hole spacetimes (\cf
  Figure~\ref{fig:bbh_constraints}).}
\label{fig:gw_infnorm}
\end{center}
\end{figure}

\begin{table*}
 \begin{tabular}{|l|c|ccccc|ccc|}
\hline
\hline
outer boundary & $h_0/M$   & $N_\mathrm{ang}$ & $R_\mathrm{in}/M$ & $R_\mathrm{out}/M$ & $N_\mathrm{lev.}$ & $r_{\text{lev}}/M$ & $(1 - \phi_{\rm Z4d} / \phi_{\rm B})$ & $(1 - \phi_{\rm Z4u} / \phi_{\rm B})$ & $(1 - \phi_{\rm Z4d} / \phi_{\rm Z4u})$ \\
\hline
multiblock, caus. discon. & $0.80$~~  &~~$33$~~  & $40.00$  & $2192.80$ & $6$ &~~$(12, 6, 3, 1.5, 0.6)$ &~~$0.0445$~~  &  $0.0465$   & $0.00230$\\
multiblock, caus. discon. & $0.60$~~  &~~$43$~~  & $39.60$  & $2192.40$ & $6$ &~~$(12, 6, 3, 1.5, 0.6)$ &~~$0.0315$~~  &  $0.0335$   & $0.00175$\\
multiblock, caus. discon. & $0.48$~~  &~~$53$~~  & $39.84$  & $2192.16$ & $6$ &~~$(12, 6, 3, 1.5, 0.6)$ &~~$0.0245$~~  &  $0.0255$   & $0.00135$\\
multiblock, caus. discon. & $0.40$~~  &~~$65$~~  & $40.00$  & $2192.40$ & $6$ &~~$(12, 6, 3, 1.5, 0.6)$ &~~$-$~~  &  $-$   & $-$\\
\hline
multiblock, caus. con.    & $0.60$~~  &~~$43$~~  & $39.60$  & $350.40$  & $6$ &~~$(12, 6, 3, 1.5, 0.6)$ &~~$-$~~  &  $-$   & $-$\\
\hline
Cartesian,  caus. con.    & $1.20$~~  &~~$0$~~   & $-$  & $199.20$ & $7$&~~$(110, 12, 6, 3 , 1.5, 0.6)$ &~~$-$~~  &  $-$   & $-$\\
\hline
\hline
\end{tabular}
\caption{Properties of the black-hole binaries simulated. The first
  column indicates the type of outer boundary and whether causally
  connected. $h_0$ is the grid spacing on the coarsest Cartesian grid,
  which is equal in all cases to the radial grid spacing in the
  angular patches. $N_\mathrm{ang}$ is the number of cells in the
  angular directions in the angular patches. $R_\mathrm{in}$ and
  $R_\mathrm{out}$ are the inner and outer radii of the angular
  patches. $N_\mathrm{lev.}$ is the number of refinement levels
  (including the coarsest) on the Cartesian grid, and
  $2\,r_{\text{lev}}$ indicates the size of the cubical refinement
  boxes centered on each black hole. The unit of the spacetime mass
  $M$ is chosen such that each black hole has mass $0.5 M$ in both the
  single and binary black cases. Finally, the last three columns
  contain the relative difference in the $\ell=m=2$ gravitational-wave
  phase between evolutions carried out with either the BSSNOK
  formulation ($\phi_{\rm B}$), the CCZ4 formulation with damping
  terms ($\phi_{\rm Z4d}$), or the CCZ4 formulation without damping
  terms ($\phi_{\rm Z4u}$).}
\label{tab:grid}
\end{table*}

The solution of the pulse at any time is just given by the advection
of the initial profile of the gauge wave, which can be set to be
smooth and periodic by choosing a sine-like initial data of the
type~\cite{Alcubierre:2003pc}
\begin{equation}
\label{Hsinus}
    h(x,t=0) = 1 - A~\sin\left(\frac{2\pi x}{L}\right)\,,
\end{equation}
with an amplitude $A<1$. Although this test is
apparently trivial as it does not involve the solution of the Einstein
equations in a very nonlinear regime, it nevertheless represents a
serious benchmark even for formulations as robust as BSSNOK, which
indeed does not pass it~\cite{Babiuc:2007vr}.

Following~\cite{Babiuc:2007vr}, we choose an amplitude of $A=0.1$ in a
domain of $L=1$ with three uniform resolutions $h_0/L = \{1/50, 1/100,
1/200\}$ and periodic boundary conditions. Notice that the metric form
(\ref{Minktrans}) corresponds to an harmonic slicing condition with
zero shift, so we have to change our preferred coordinate choice (\ie
the $1+\log$ slicing with the Gamma-driver) to perform this
test. Furthermore, different implementations of the CCZ4 formulation:
one in which the constraints are \textit{damped} with coefficients
$\kappa_1=1/L$ and $\kappa_2=0$, and one in which the constraints are
\textit{undamped}, \ie $\kappa_1=0=\kappa_2$. We will refer to these
two cases as to ``Z4d'' and ``Z4u'', respectively (Note that in these
tests the shift is set to zero and hence we do not need to specify a
value for $\kappa_3$, which we take to be one).

The infinity-norm of the Hamiltonian constraint relative to
simulations at the highest resolution is displayed in
Fig.~\ref{fig:gw_infnorm} for the damped CCZ4 formulation (black solid
line), for the undamped CCZ4 formulation (blue dashed line), and for
the BSSNOK formulation (red dotted line). Clearly, the BSSNOK and the
CCZ4 formulation without damping terms fail before $50$ crossing times
(BSSNOK after $42$ crossing times and Z4u after $56$ crossing times)
as indicated by the an exponential increase in the violation of the
energy constraint.  However, with the addition of the damping terms,
the CCZ4 formulation is able to accurately evolve this solution for
more than $1000$ crossing times, while preserving the profile of the
pulse. Furthermore, we have verified that the evolved solution
converges to the expected spatial-discretization order (\ie either 4th
or 8th order), with only a very small phase error when using the 8th
order scheme.

Overall, this test shows that the dynamical control of the energy
constraint via the damping term $\kappa_1$ is \textit{crucial} to
attain a stable evolution, even in such a simple type of
spacetimes. We also note that this test is more demanding for
conformal formulations, where there is more than one component of the
metric which is nontrivial. This is confirmed by comparing our results
with those in Ref.~\cite{Alic:2009}, where the standard Z4
formulation, \ie not implementing a conformal decomposition, was able
to pass this test without the need of damping terms. The GH
formulation also passes this test.


\subsection{Black-Hole Spacetimes}
\label{Black-Hole Spacetimes}

Before considering black-hole binaries, we have tested extensively our
new CCZ4 formulation in the evolution of single nonspinning black
holes. This has allowed us to determine how different choices for the
damping coefficients $\kappa_1$ and $\kappa_2$ influence the solution
and, in particular, the violation of both the ADM and the $Z_{\mu}$
constraints. In this way we have concluded that most of the dynamics
in the evolution of the constraint equations comes from the first
damping coefficient, so that $\kappa_2=0$ represents a sensible choice
and is the one that we will consider hereafter. On the other hand,
increasing values of $\kappa_1$ produce lower violations of the
constraints and a value of $\kappa_1 \approx 0.1/M$ seems optimal in
this sense. Higher values, in fact, lead only to marginal improvements
of the solution, but also tend to increase the stiffness of the
damping terms.

\begin{figure}
\begin{center}
\includegraphics[width=8.0cm]{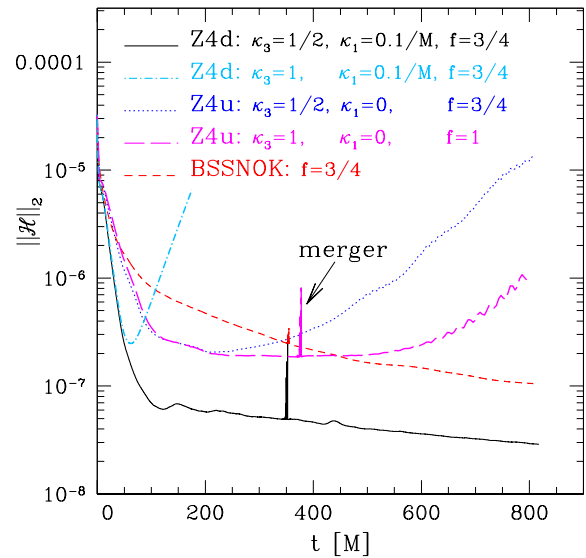}
\caption{L2-norm of the Hamiltonian constraint for the noncovariant CCZ4
  formulation with and without damping terms (black solid line and
  blue dotted line, respectively), for the covariant CCZ4 formulation 
  with and without damping terms
  (light-blue dot-dashed line and magenta long-dashed line,
  respectively), 
  and for the BSSNOK formulation (red
  dashed line). Also indicated are the different values of coefficient
  $f$ in the shift Eq.~\eqref{gammadriver1}, which however do not
  introduce qualitatively different behaviors. The data refers to the
  a simulations having a coarse resolution of $h_0/M=0.48$ and outer
  boundary at $R_{\rm out}=2192.16\,M$. 
}
\label{fig:bbh_constraints}
\end{center}
\end{figure}

An important and unexpected result obtained when implementing the CCZ4
formulation in black-hole spacetimes is that subtle and nonlinear
couplings can occur, leading to unstable evolutions also for those
choices of the coefficients that are perfectly stable in other
spacetimes. While, in fact, we have carried out stable evolutions of
the robust-stability test with the covariant and damped CCZ4
formulation (\ie with $\kappa_3=1$ and $\kappa_1 \neq 0$), we were not
able to obtain stable evolutions of black-hole spacetimes with
$\kappa_3=1$, although the growth time of the instability does change
with the values of $\kappa_1$ (see discussion around
Fig.~\ref{fig:bbh_constraints}). Clearly, nontrivial couplings seem
to appear between these coefficients, which depend on the degree of
nonlinearity and which deserve further investigation to be properly
understood.

On the whole, and as we will detail below, we have found that
\emph{accurate} and \emph{stable} evolutions of binary black-hole
spacetimes can be obtained with the damped noncovariant Z4 systems
(\ie with \hbox{$\kappa_3=1/2$}, $\kappa_1 =0.1/M$). On the other
hand, covariant and conformal Z4 formulations that are either damped
(\ie with $\kappa_3=1$, $\kappa_1\neq 0$), or undamped (\ie with
$\kappa_3=1$, $\kappa_1=0$), have been found to lead to
\emph{unstable} evolutions, although on rather different timescales
and with variable degree of accuracy (see discussion below).

The initial data of the binary black-hole evolutions is obtained from
a circular-orbit condition at the third post-Newtonian
order~\cite{Husa:2007rh} and corresponds to an equal-mass nonspinning
binary with an initial coordinate separation of $D=8\,M$. The binary
performs about $3.5$ orbits before merging and settles to an isolated
spinning black hole after $t \approx 360\,M$. To carry out a
meaningful comparison, the binary is evolved with the BSSNOK and the
CCZ4 formulations keeping the same choice for the gauges, namely the
$1+\log$ slicing condition and the Gamma-driver shift condition with
$f = {3}/{4}$, $\eta=2/M$, and the same grid setup. For the latter, in
particular, we have considered three different choices aimed at
determining the influence of the outer boundaries on the quality of
the solution. This is a point discussed in
Refs.~\cite{Bernuzzi:2009ex, Ruiz:2010qj}, where it was argued that
the Z4c formulation is more sensitive than the BSSNOK one to incorrect
(or constraint-violating) boundary conditions. As a result, we consider
three different classes of simulations depending on the treatment of
the outer boundary: (i) multiblock padding and spherical outer
boundary which is causally disconnected (\ie~at $\sim 2200\,M$ for a
simulation lasting $\sim 800\,M$); (ii) multiblock padding and
spherical outer boundary which is causally connected (\ie~at $\sim
350\,M$); (iii) Cartesian outer boundary which is causally connected
(\ie~at $\sim 200\,M$). For case (i), we reduce the order of the
finite-difference operator at the outer boundary but, because it is
causally disconnected, the initial conditions are preserved there. For
case (ii), instead, we impose reflecting boundary conditions so as to
``stress'' the solution with data from the outer boundary which is
constraint-violating and injected mostly at the time of the
reflection. Finally, in case (iii) we have applied ordinary, outgoing
Sommerfeld boundary conditions to all variables, again triggering
violations in the constraint equations.

All the properties of the grid structure and the treatment of the
outer boundary are summarized in Table~\ref{tab:grid}, where $h_0$ is
the grid spacing on the coarsest Cartesian grid, which is equal in all
cases to the radial grid spacing in the angular
patches. $N_\mathrm{ang}$ is the number of cells in the angular
directions in the angular patches, while $R_\mathrm{in}$ and
$R_\mathrm{out}$ are the inner and outer radii of the angular patches,
respectively. In the case of a Cartesian outer boundary, $R_{\rm out}$
represents the distance to the outer boundary along coordinate
lines. Finally, $N_\mathrm{lev.}$ is the number of refinement levels
(including the coarsest) on the Cartesian grid, while
$2\,r_{\text{lev}}$ indicates the size of the cubical refinement boxes
centered on each black hole.

As final remark before discussing the results, we note that all the
rest being the same, at any given resolution the CCZ4 system has a
smaller violation of the constraints than the BSSNOK one. At the same
time, however, because the violations of both the energy and momentum
constraints are part of the evolution equations in the CCZ4 system,
the latter is more strongly affected than BSSNOK one, for which only
the violations of the momentum constraint are included in the
evolution system. As a result, the CCZ4 formulation requires a
comparatively higher minimum-resolution treshold in order to enter a
convergent regime.

\begin{figure}
\begin{center}
\includegraphics[width=8.0cm]{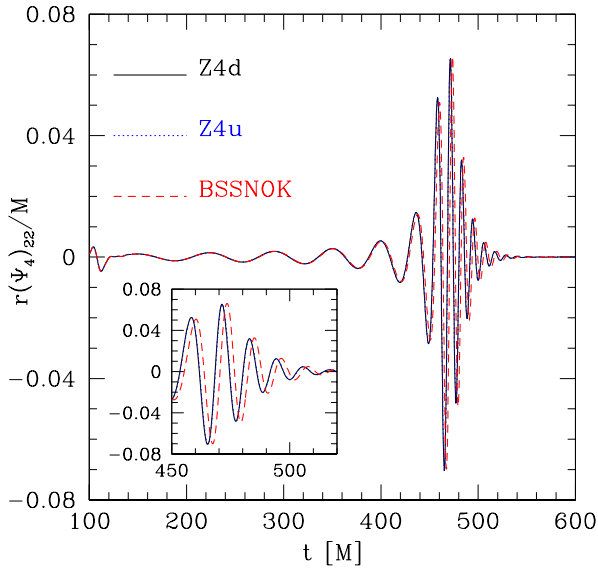}
\caption{Real part of the $\ell=m=2$ mode of the gravitational
  waveform $\Psi_4$ for an equal-mass nonspinning black-hole
  binary. Different lines refer to evolutions with the noncovariant
  formulation with and without damping terms, \ie with
  \hbox{$\kappa_3=1/2$} and $\kappa_1 =0.1/M$, $\kappa_2=0$ (Z4d), or
  \hbox{$\kappa_3=1/2$} and $\kappa_1= \kappa_2=0$ (Z4u). The two
  evolutions are indicated, respectively, as Z4d and with a black solid
  line or as Z4u and with a blue dotted line; the BSSNOK formulation
  is shown with a red dashed line. Shown in the inset is a
  magnification of the merger.}
\label{fig:bbh_waveform}
\end{center}
\end{figure}

A first comparison of the behavior of the different formulations is
offered in Fig.~\ref{fig:bbh_waveform}, where we show the $\ell=m=2$
mode of the gravitational waveform $\Psi_4$ as extracted on a sphere
of coordinate radius $r=100\,M$ (see~\cite{Pollney:2009yz} for details
on the extraction procedure). Different lines refer to simulations
using either the noncovariant formulation with damping terms, \ie
with \hbox{$\kappa_3=1/2$} and $\kappa_1 =0.1/M$, $\kappa_2=0$ (Z4d,
black solid line), or to the noncovariant formulation without damping
terms, \ie with \hbox{$\kappa_3=1/2$} and $\kappa_1= \kappa_2=0$ (Z4u,
blue dotted line). Also shown as a reference is a simulation with the
BSSNOK formulation (red dashed line) using the same numerical
setup. The simulations refer to the highest resolution (\ie
$h_0/M=0.48$) and the grid having the multiblock padding and an outer
boundary at $R_{\rm out}=2192.16\,M$.

The first obvious thing to note is that all simulations lead to a
stable merger and ringdown at all the resolutions
considered. Furthermore, while a small phase difference is present
between the Z4 and the BSSNOK runs, this difference is very small and
$\Delta \phi \lesssim 0.02$ rad over the whole simulation. As a
comparison, the phase difference between the Z4 and the Z4u
simulations is $\Delta \phi \lesssim 0.002$ rad (see
Table~\ref{tab:grid} for the relative maximum differences).

\begin{figure}
\begin{center}
\includegraphics[width=8.0cm]{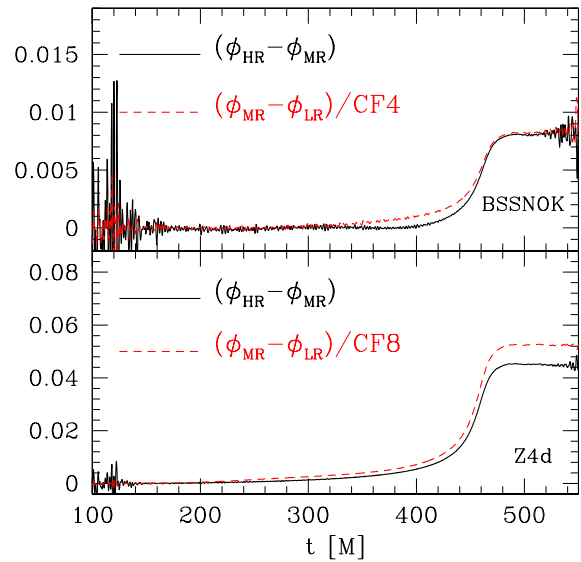}
\caption{Differences in the phase evolutions at the high, medium and
  low resolutions, respectively (these are indicated as HR, MR
  and LR). The top panel refers to the BSSNOK formulation, while
  the bottom one the the noncovariant damped CCZ4 formulation
  (Z4d).  The differences between the low and medium resolutions are
  also scaled with the appropriate convergence coefficients (marked as
  ${\rm CF4}$ and ${\rm CF8}$, see text) to highlight the convergence
  order of the solution; all the data refers to simulations with a
  multiblock padding and causally disconnected outer boundary. Note
  that at these resolutions the CCZ4 formulation has larger phase
  errors, but due its higher convergence factor, these errors are
  expected to decay at a faster rate than for BSSNOK. }
\label{fig:bbh_convergence}
\end{center}
\end{figure}

Although the phase differences between the waveforms obtained with the
two formulations is relatively small, it also decreases with the
resolution, thus indicating that both formulations would yield the
same phase evolution in the continuum limit. The rate of convergence,
however, is different when considering either the BSSNOK or the CCZ4
formulation. This is shown in Fig.~\ref{fig:bbh_convergence}, where we
report the residuals in the phase evolutions at the high, medium and
low resolutions, respectively (these are indicated as ``HR'', ``MR''
and ``LR''). The differences between the low and medium resolutions
are also scaled to highlight the convergence order of the
solution. More specifically, the HR, MR and LR refer to
simulations with the coarsest resolutions of $h_0/M=0.6, 0.48, 0.4$
(\cf Table~\ref{tab:grid}). The convergence coefficients corresponding
to these resolutions and used for rescaling are ${\rm CF4}=3.0898$,
for a convergence factor of $4.5$ in the BSSNOK case, and ${\rm
  CF8}=7.1906$ for a convergence factor of $8.5$ in the Z4d case.
Note however that, as mentioned above, the CCZ4 formulation needs a
higher resolution to enter the convergence regime, while a triplet of
resolutions with $h_0/M=0.8, 0.6, 0.48$ would be enough to show
convergence at about 4th order for the BSSNOK runs.

Beside this minimum resolution threshold, the additional computational
expenses required by the CCZ4 formulations are not significant. The
difference with the BSSNOK system consists in an additional evolution
equation for the scalar variable $\Theta$, which would amount to
solving 25 evolution equations (instead of 24 as in BSSN), implying
around $4\%$ higher computational costs. However this is an
over-estimate, as in reality the time spent in computing the evolution
equations depends on the computational infrastructure. In our case, it
is about half of the total time of a binary black-hole simulation,
while the other half is dedicated to mesh-refinement,
gravitational-wave extraction and other analysis routines.

All in all, we find that for the highest resolutions used the results
of the BSSNOK runs converge at about 4th order (top panel in
Fig.~\ref{fig:bbh_convergence}), while the Z4d runs converge at about
8th order (bottom panel in Fig.~\ref{fig:bbh_convergence}); in both
cases, the convergence order is lost in the very final stages of the
merger. It is a present unclear why the two formulations yield, with
the same computational infrastructure, two different convergence
rates. It is possible that the constraint-damping properties of the
CCZ4 formulation are able to suppress the small violations coming from
the reflections across refinement boundaries, that are a major source
of error and one of the largest obstacles to attain clean
convergence. However, more efforts (and considerable computational
costs) need to be invested to assess whether this is the correct
explanation.

A useful way to appreciate the different behavior of the two
formulations is shown in Fig.~\ref{fig:bbh_constraints}, which reports
the evolution of the L2-norm of the ADM energy (\ie the violation of
the Hamiltonian constraint) for the covariant CCZ4 formulation with
and without damping (light-blue dot-dashed line and magenta
long-dashed line, respectively), for the noncovariant CCZ4
formulation with and without damping (black solid line and blue dotted
line, respectively), and for the BSSNOK formulation (red dashed
line). We also report the different values of coefficient $f$ in the
shift Eq.~\eqref{gammadriver1}, which does change the growth rate
of the unstable simulations, but does not remove the instability in
the case of the fully covariant formulation\footnote{We have performed
  simulations also with $\kappa_3=1, \kappa_1=0.1/M, f=1$, or
  $\kappa_3=1, \kappa_1=0.1/M, f=3/4$, and $\kappa_3=1, \kappa_1=0,
  f=3/4$; in all cases we have found an instability (although with
  different growth rates), which we do not report to avoid overloading
  Fig.~\ref{fig:bbh_constraints}.}. The data refers to simulations
having a coarse resolution of $h_0/M=0.48$ and outer boundary placed
at $R_{\rm out}=2192.16\,M$, but similar behaviors have been seen
also at higher and lower resolutions.

Note that as the initial data settles and the evolution proceeds, the
CCZ4 formulation shows a violation of the Hamiltonian constraint
smaller than for the BSSNOK case (the L2-norm being at least 1 order
of magnitude smaller), hence yielding a more accurate solution of the
Einstein equations. However, after this initial stage, the evolutions
with the CCZ4 formulation can be considerably different according to
the choice made for the parameters $\kappa_3$ and $\kappa_1$. More
specifically, the covariant and damped system (\ie $\kappa_3=1$,
$\kappa_1\neq0$) exhibits a very rapid violation of the constraint at
$\sim 100\,M$ and inevitably leads to a code crash (light-blue
dot-dashed line in Fig.~\ref{fig:bbh_constraints}). Other variants of
the CCZ4 formulation, on the other hand, show a different
behavior. In particular, both of the undamped CCZ4 formulations (\ie
$\kappa_3=1/2, 1$, $\kappa_1=0$) lead to a successful merger, which
can be easily identifiable as the peak at about $\simeq 350-380\,M$,
and which is due to larger local violations of the constraints as the
merger takes place\footnote{Note that the time of merger is a gauge
  dependent quantity and can therefore take place at slightly
  different times in different formulations.}. At the same time,
however, both implementations show a growth of the constraint
violation (blue dotted line and magenta long-dashed line). This growth
can be rather slow in the case $f=1$, but it is likely to yield
unstable evolutions on very long timescales. Finally,
Fig.~\ref{fig:bbh_constraints} shows that a noncovariant and damped
implementation of the CCZ4 formulation (\ie $\kappa_3=1/2$,
$\kappa_1\neq 0$; black solid line) leads not only to a stable merger
and subsequent evolution, but it also provides a violation of the
constraints which is at least 1 order of magnitude smaller than the
corresponding one obtained with the BSSNOK evolution (red dashed
line). This is one the main results of this paper and the ultimate
justification for investigating this new formulation of the equations.

We note that the behavior of the constraints described above for the
CCZ4 formulation is indeed very similar to what already experienced by
many groups implementing the GH formulation\footnote{We recall that
  GH formulation can can be seen as a reduction of the Z4 formalism
  \cite{Bona:2003fj}.}. In that case, in fact, the addition of the
damping terms was crucial to achieve stable black-hole
evolutions~\cite{Pretorius:2005gq, Pretorius:2006tp,
  Szilagyi:2006qy}. Altogether, the evolution shown in
Fig.~\ref{fig:bbh_constraints} already provides the needed evidence
that the new CCZ4 formulation, once suitable damping terms are added
and the boundary conditions do not play a role, represents a
considerable improvement over the standard BSSNOK formulation. In what
follows we will show that this continues to be the case also when the
outer boundaries are chosen to produce incorrect data, or when they
are placed very close to the merging binary.

\begin{figure}
\begin{center}
\includegraphics[width=8.0cm]{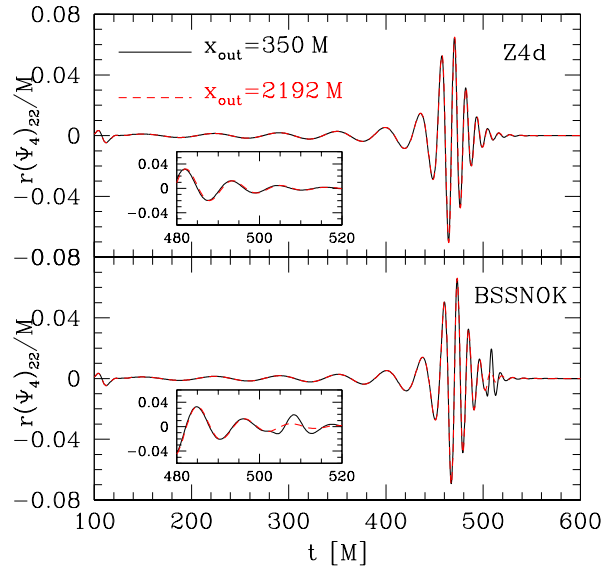}
\caption{Real part of the $\ell=m=2$ mode of the gravitational
  waveform $\Psi_4$ (black solid line) extracted at $r=100\,M$ for
  simulations having a coarse resolution $h_0/M=0.60$ and an outer
  boundary which is causally connected and at $R_{\rm
    out}=350.40\,M$. The top panel refers to a simulation using the
  noncovariant and damped implementation of the CCZ4 formulation (\ie
  $\kappa_3=1/2$, $\kappa_1\neq 0 $), while the bottom one to a
  simulation using the BSSNOK formulation; also shown are the
  corresponding waveforms obtained when $R_{\rm out}=2192.40\,M$ (red
  dashed lines).}
\label{fig:gws_boundary}
\end{center}
\end{figure}

Figure~\ref{fig:gws_boundary} reports with black solid lines the
$\ell=m=2$ mode of the gravitational waveform $\Psi_4$ extracted at
$r=100\,M$ for simulations having a coarse resolution $h_0/M=0.60$ and
an outer boundary which is causally connected and at $R_{\rm
  out}=350.40\,M$ (\cf Table~\ref{tab:grid}). The top panel, in
particular, refers to a simulation using the noncovariant and damped
implementation of the CCZ4 formulation (\ie Z4d, with $\kappa_3=1/2$,
$\kappa_1\neq 0 $), while the bottom one to a simulation using the
BSSNOK formulation. Also shown with red dashed lines are the
corresponding waveforms obtained when the outer boundary is causally
disconnected and at $R_{\rm out}=2192.40\,M$. As shown more clearly in
the two insets, the CCZ4 formulation yields waveforms which are
essentially identical and are unaffected by the constraint-violating
outer boundaries. This is to be contrasted with the evolution
performed with the BSSNOK formulation and which shows strong signs of
reflection at $t\simeq 510\,M$.

\begin{figure}
\begin{center}
\includegraphics[width=8.0cm]{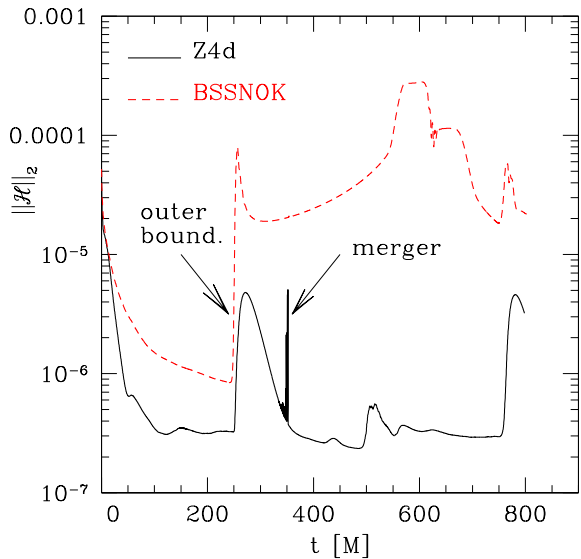}
\caption{L2-norm of the Hamiltonian constraint for the noncovariant
  and damped implementation of the CCZ4 formulation (\ie Z4d with
  $\kappa_3=1/2$, $\kappa_1\neq 0 $), and for the BSSNOK formulation
  (red dashed line). The data refers to the a simulations having a
  coarse resolution of $h_0/M=0.60$ and outer boundary placed at
  $R_{\rm out}=350.40\,M$.}
\label{fig:HC_rout_350}
\end{center}
\end{figure}

The reason behind this different behavior is to be found in the
different way in which the two formulations handle the
constraint-violations coming from the outer boundaries and is best
appreciated in Fig.~\ref{fig:HC_rout_350}, where we show again the
L2-norm of the ADM energy for the noncovariant and damped
implementation of the CCZ4 formulation (\ie Z4d with $\kappa_3=1/2$,
$\kappa_1\neq 0 $) and for the BSSNOK formulation. Note that both
suffer of a very large increase at $t\sim 250\,M$ when the waves from
the initial gauge settling of the binary, propagating at a speed of
$v_{\rm g} \sim \sqrt{2}$, reach the outer boundary at $R_{\rm
  out}=350.40\,M$ and lead to larger violations. Also note that this
increase in the constraint violation happens much earlier than the one
associates with the merger (which is at $t \sim 350\,M$). As evident
from Fig.~\ref{fig:HC_rout_350}, the CCZ4 is able to recover
efficiently from this violation, and the damping terms act in such a
way that by $t\sim 400\,M$ the violation is completely removed, with
the Hamiltonian constraint brought back to its minimum value. By
contrast, the evolution with the BSSNOK formulation never recovers
from the boundary contamination, leading to an increasing violation
responsible for the incorrect behavior discussed in
Fig.~\ref{fig:gws_boundary}. The CZZ4 formulation experiences another
increase in the violation at $t\sim 750\,M$, when the gauge waves
coming from the binary reach again the outer boundary, but once again
the constraint damping terms act so as to remove the violation.

\begin{figure}
\begin{center}
\includegraphics[width=8.0cm]{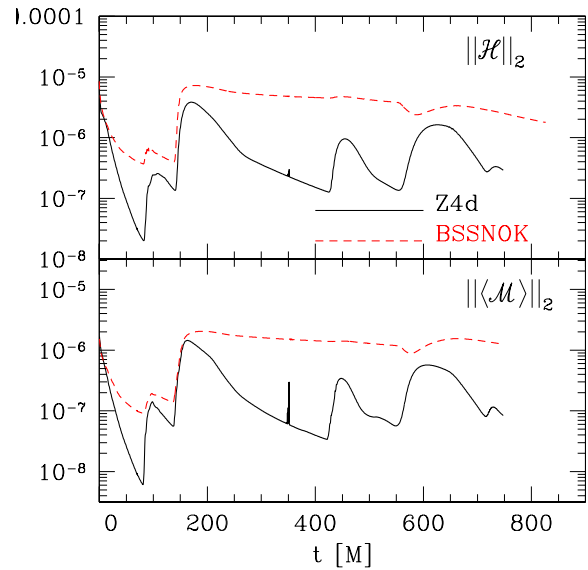}
\caption{L2-norm of the Hamiltonian constraint (top panel) and of the
  root-mean-square of the momentum constraint (bottom panel) for the
  noncovariant and damped implementation of the CCZ4 formulation (\ie
  Z4d with $\kappa_3=1/2$, $\kappa_1\neq 0 $, black solid lines), and
  for the BSSNOK formulation (red dashed lines). The data refers to
  simulations having a coarse resolution of $h_0/M=1.20$ and
  outer boundary at $R_{\rm out}=199.20\,M$.}
\label{fig:HC_rout_200}
\end{center}
\end{figure}

An additional and concluding evidence of the constraint-damping
properties of the CCZ4 formulation is shown is
Fig.~\ref{fig:HC_rout_200}, where we report the evolution of the
L2-norm of the Hamiltonian constraint (top panel) and of the
root-mean-square of the momentum constraint (bottom panel) for the
noncovariant and damped implementation of the CCZ4 formulation (\ie
Z4d with $\kappa_3=1/2$, $\kappa_1\neq 0 $, black solid lines), and
for the BSSNOK formulation (red dashed lines). The data refers to
simulations performed with a plain Cartesian outer boundary which is
very close to the binary and at $R_{\rm out}=199.20\,M$ (\cf
Table~\ref{tab:grid}). As in the previous figure, also here it is
possible to detect the increase of the constraint violations when
gauge waves from the binary have reached the outer boundary at $t\sim
140\,M$.

Also in this case, the damping terms in the equations remove
rapidly the violations, which decay exponentially to their minimum
values. Because the boundary is so close-in, this behavior of rapid
increase and exponential decay takes place at least 3 times, both
for the Hamiltonian and momentum constraints. Any formulation of the
Einstein equations having this type of behavior is obviously
preferable over one in which the violations are trapped in the
computational domain and are not allowed to be damped.

\section{Conclusions}
\label{conclusions}


By starting from the Z4 formulation~\cite{Bona:2003fj} and by
including \textit{all} the nonprincipal terms coming from the
covariant form of the equations, we have introduced the CCZ4
formulation, \ie the conformal and covariant formulation of the Z4
system, and proposed it as a new and effective way to solve
numerically the Einstein equations in arbitrary spacetimes.

The new set of equations combines the most important features of the
commonly used formulations of the Einstein equations employed in
numerical-relativity calculations. In particular, it is able to make
use of well-tested and robust gauge conditions which remove the need
of excision and, at the same time, it is able to control dynamically
the violation of the constraint equations and to rapidly suppress them
when they occur. 

We have validated the robustness of the CCZ4 evolution system by
performing a number of tests both in flat and in black-hole
spacetimes. We have thus found that the CCZ4 formulation without
damping terms does not pass the standard gauge-advection test, in
analogy with the behavior of the BSSNOK formulation. However, when
the damping terms are switched on, the new CCZ4 formulation passes the
test stably and accurately. 

This ability of the formulation to control and damp violations in the
constraint equations has been confirmed also through the simulation of
nonspinning black-hole binaries, which have been followed for about
three orbits before merging to a rapidly rotating black hole. Through
a series of simulations at different resolutions and with different
treatments of the outer boundary -- handled either with multiblocks
and placed at a causally-disconnected distance, or with a Cartesian
box and placed close to the binary -- we have shown that \emph{not all}
of the implementations of the CCZ4 formulation lead to stable
evolutions of binary black-hole spacetimes. 

Rather, we have found that the covariant form of the CCZ4 formulation,
in conjunction with the use of damping terms, leads to exponentially
growing modes that rapidly destroy the numerical
solution. Fortunately, the use of a noncovariant formulation and of
damping terms leads not only to a stable evolution, but it also
provides a violation of the constraints which is at least 1 order of
magnitude smaller than the corresponding one obtained with the BSSNOK
evolution. A close comparison with simulations performed with the
BSSNOK formulation using the same numerical setup, has also revealed
that the CCZ4 formulation can efficiently recover from large
violations of the constraints, with the damping terms rapidly removing
constraint violations produced at the outer boundary. By contrast,
evolutions with the BSSNOK formulation experiencing similar violations
never recover from the boundary contamination, leading to an
increasing violation and incorrect gravitational waves.

Because the changes necessary to implement the new conformal
formulation in BSSNOK codes and the additional computational costs are
very small, we propose the new formulation as a new standard for the
numerical solution of the Einstein equations in generic 3D
spacetimes. We expect, in fact, that a numerical solution of the
Einstein equations having smaller violations of the constraints will
also yield a more accurate modelling of the gravitational-wave
emission, both in vacuum and nonvacuum spacetimes.

At the same time, however, much remains to be done to fully understand
the role played by the damping coefficients in fully nonlinear regimes
and in the covariant form of the CCZ4 formulation. Our experience with
binary black-hole spacetimes has revealed, in fact, that there are
situations in which the damping of the constraints interferes
negatively with a fully covariant form of the CCZ4 formulation,
leading to unstable evolutions. In these cases, even small changes in
the covariant character of the equations (\eg by using $\kappa_3=0.9$
instead of $\kappa_3=1$) allows one to use nonzero damping
coefficients and hence to obtain a smaller violation of the
constraints. A systematic investigation of the space of parameters
$\kappa_1\times\kappa_2\times\kappa_3$ is difficult due to the large
computational costs of these simulations, but is clearly needed for a
deeper understanding of the behavior of the CCZ4 formulation. Much of
our future work will be dedicated to elucidate this point.

\bigskip


\acknowledgments

We thank Ian Hinder and Barry Wardell for the analysis tools used in
this work, Jose-Luis Jaramillo for useful discussions on the conformal
formalism, Carlos Lousto for comparison with his implementation of the
CCZ4 formulation, and Sebastiano Bernuzzi, David Hilditch and Milton
Ruiz for discussions on their Z4c formulation. Partial support comes
from the European Union FEDER funds, by the Spanish Ministry of
Science and Education (projects FPA2010-16495 and CSD2007-00042), by
the DFG Grant SFB/Transregio 7 and by the NSF grant PHY-0803629. The
computations were performed at the AEI and on the Teragrid network
(allocation TG-MCA02N014).



%
%

\bibliographystyle{apsrev-nourl-noeprint}
\bibliography{aeireferences}

\end{document}